\documentclass[prd,a4paper,nofootinbib,showpacs]{revtex4}
\usepackage{bm}
\usepackage{amsfonts}
\begin{document}
%-------------------------------------------------------------------------
% Definition needed for the heading
%-------------------------------------------------------------------------
\def\Barcelo{Barcel\'o}
\def\Schrodinger{Schr\"odinger}
%-------------------------------------------------------------------------
\title[Stacking a 4D geometry into ...]{Stacking
a 4D geometry into an Einstein-Gauss-Bonnet bulk}
%-------------------------------------------------------------------------
\author{Carlos \Barcelo}
%\email{carlos.barcelo@port.ac.uk}
%\homepage{}
%\thanks{Supported by the EC under contract HPMF-CT-2001-01203}
%\altaffiliation{}
\affiliation{Institute of Cosmology and Gravitation,
University of Portsmouth, Portsmouth PO1 2EG, Britain}
%-------------------------------------------------------------------------
\author{Roy Maartens}
%\email{roy.maartens@port.ac.uk}
%\homepage{}
%\thanks{}
%\altaffiliation{}
\affiliation{Institute of Cosmology and Gravitation,
University of Portsmouth, Portsmouth PO1 2EG, Britain}
%-------------------------------------------------------------------------
\author{Carlos F. Sopuerta}
%\email{carlos.sopuerta@port.ac.uk}
%\homepage{}
%\thanks{}
%\altaffiliation{}
\affiliation{Institute of Cosmology and Gravitation,
University of Portsmouth, Portsmouth PO1 2EG, Britain}
%-------------------------------------------------------------------------
\author{Ferm\'{\i}n Viniegra}
%\email{fermin.viniegra@port.ac.uk}
%\homepage{}
%\thanks{}
%\altaffiliation{}
\affiliation{Institute of Cosmology and Gravitation,
University of Portsmouth, Portsmouth PO1 2EG, Britain}
%-------------------------------------------------------------------------

%-------------------------------------------------------------------------
%\date{16 October 2002; gr-qc/0211013; \LaTeX-ed \today}
%-------------------------------------------------------------------------
\begin{abstract}
%-------------------------------------------------------------------------
In Einstein gravity there is a simple procedure to build
$D$-dimensional spacetimes starting from $(D-1)$-dimensional ones,
by stacking any $(D-1)$-dimensional Ricci-flat
metric into the extra-dimension. We analyze this
procedure in the context of Einstein-Gauss-Bonnet gravity, and find
that it can only be applied to metrics with a constant Krestschmann
scalar.  For instance, we show that solutions of the black-string type
are not allowed in this framework.

%-------------------------------------------------------------------------
\end{abstract}
%-------------------------------------------------------------------------
%-------------------------------------------------------------------------
\maketitle
%-------------------------------------------------------------------------
%-------------------------------------------------------------------------
% Author-specific definitions
%-------------------------------------------------------------------------
\def\half{{1\over 2}}
\def\L{{\mathcal L}}
\def\S{{\mathcal S}}
\def\d{{\mathrm{d}}}
\def\x{{\mathbf x}}
\def\v{{\mathbf v}}
\def\im{{\rm i}}
\def\etal{{\emph{et al\/}}}
\def\det{{\mathrm{det}}}
\def\tr{{\mathrm{tr}}}
\def\ie{{\emph{i.e.}}}
\def\bnabla{\mbox{\boldmath$\nabla$}}
\def\Box{\kern0.5pt{\lower0.1pt\vbox{\hrule height.5pt width 6.8pt
    \hbox{\vrule width.5pt height6pt \kern6pt \vrule width.3pt}
    \hrule height.3pt width 6.8pt} }\kern1.5pt}
\def\HRULE{{\bigskip\hrule\bigskip}}
%-------------------------------------------------------------------------

%---------------------------------------------------------------------
\section{Introduction}
\label{S:introduccion}
%---------------------------------------------------------------------

One of the main ingredients in most attempts to unify the four
fundamental interactions known at present in nature is the existence
of additional dimensions.  If there is a limit in which one can make
sense of these additional dimensions classically, it will be
fundamentally important to know what the field equations are that govern the
higher-dimensional spacetime.  Lovelock~\cite{Lovelock:1971yv} proved
that for an arbitrary number of dimensions, the most general classical
gravitational Lagrangian with associated field equations ${\cal
G}_{AB}=0$, such that ${\cal G}_{AB}$ is symmetric, divergence-free, 
and constructed with up to second derivatives of the metric,
is formed by a linear combination (with arbitrary coefficients) of the
dimensionally-extended Euler densities. As
is well known, in four dimensions there are only two Euler densities
that are not topological invariants, and therefore have a non-trivial
dynamical content. They are the scalar curvature term and the
cosmological constant term in the Einstein-Hilbert action.
Experimentally it has been possible to determine the values of their
two associated coefficients: Newton's constant and cosmological
constant.  In $D=5$ and beyond there are additional Euler densities
that have to be considered. The relative weight of these additional
terms in the dynamics of the system is something to be determined
experimentally.

From our 4-dimensional point of view, the higher-dimensional geometries
of more immediate interest are those suitable for a standard or exotic
Kaluza-Klein reduction.  In this paper we will consider only
5-dimensional geometries of this type. In five dimensions there is
only one additional Euler density to be considered: the Gauss-Bonnet
term.  Our main motivation for studying 5-dimensional geometries with
a Gauss-Bonnet term comes about from the exotic Kaluza-Klein reduction
realized in the Randall-Sundrum braneworld scenario~\cite{Randall:1999ee,Randall:1999vf}.  In this scenario our universe
is described as a 4-dimensional brane immersed in a
5-dimensional anti-de Sitter bulk.  When
the braneworld scenario is considered as a low-energy limit of
string/M theory it becomes even more natural to consider the effects
of the Gauss-Bonnet term~\cite{Zwiebach:1985uq,Boulware:1985wk}.

In the presence of Gauss-Bonnet modifications, it was first proved
that it is possible to obtain a vacuum geometry equivalent to that of
Randall-Sundrum apart from a redefinition of the
constants~\cite{Kim:1999dq}.  
This redefinition is such that it allows
a Minkowski brane in an anti-de Sitter bulk even without a bulk
negative cosmological constant and a positive 
brane tension~\cite{Meissner:2000dy}.  
A massless graviton is also shown to appear
in this construction~\cite{Kim:2000pz,Meissner:2001xg}.
In the context of self tuning mechanisms for the vanishing of 
the cosmological
constant (see for example ~\cite{Csaki:2000wz} and references therein), 
the Gauss-Bonnet term allows to avoid the presence of naked singularities 
in the bulk~\cite{Binetruy:2002ck}.
Cosmology on the brane in this scenario has received much
attention~\cite{Deruelle:2000ge,Abdesselam:2001ff,Germani:2002pt,Charmousis:2002rc,Lidsey:2002zw,Nojiri:2002hz}.
In all these works the brane is maximally
symmetric or of cosmological type.  Here, we analyze
branes with arbitrary metric (but with specific extensions into the
bulk).

In Einstein General Relativity there is an easy procedure to produce
5-dimensional solutions of the field equations by trivially extending
vacuum 4-dimensional solutions into the extra dimension~\cite{Brecher:1999xf}.  This procedure was used to find a black
string solution~\cite{Chamblin:1999by} and a plane wave solution~\cite{Chamblin:1999cj} in the realm
of braneworlds. We show that this procedure for
generating solutions does not generalize to
Einstein-Gauss-Bonnet gravity.  For the procedure to be applicable,
the 4-dimensional geometry has to satisfy an additional constraint:
its Krestschmann scalar has to be constant.  This implies in particular that
geometries of black-string type cannot be constructed in this
framework.

In the next section we review and generalize the procedure for
generating $D$-dimensional solutions starting from $(D-1)$-dimensional
ones. We show that the procedure works not only for vacuum
solutions~\cite{Brecher:1999xf}, but for Einstein manifolds. Then, in section~\ref{S:gauss-bonnet} we investigate what happens in the case of Einstein-Gauss-Bonnet
gravity with $D=5$. Finally, we discuss the results
and conclude.

%---------------------------------------------------------------------
\section{Stacking technique}
\label{S:stackingtechnique}
%---------------------------------------------------------------------

There is a set of solutions to $D$-dimensional Einstein gravity that can
be constructed in a simple way starting from vacuum $(D-1)$-dimensional
solutions.   The procedure is based on the stacking of any vacuum
$(D-1)$-dimensional solution
into the additional dimension~\cite{Brecher:1999xf}. Given any
metric $g_{\mu\nu}(x)$ such that the $(D-1)$-dimensional Einstein tensor satisfies $G_{\mu\nu}=0$, then
\begin{eqnarray}
ds^2=dy^2 + g_{\mu\nu}(x) dx^\mu dx^\nu
\end{eqnarray}
is a solution of the $D$-dimensional Einstein equations $^{(D)\!}G_{AB}=0$.
This procedure can also be used when a $D$-dimensional negative cosmological
constant $|\Lambda_D|=(D-2)(D-1)/2 l^2$ is present in the bulk: any metric
$^{(D)\!}g_{AB}$ of the form
\begin{eqnarray}
ds^2=dy^2 + e^{-2y/l} g_{\mu\nu}(x) dx^\mu dx^\nu \,,
\label{flatbrane}
\end{eqnarray}
where $g_{\mu\nu}(x)$ is a $(D-1)$-dimensional vacuum solution, is  a
solution of
\begin{eqnarray}
^{(D)\!}G_{AB}=-\Lambda_{D}\, ^{(D)\!}g_{AB}.
\label{Einstein}
\end{eqnarray}

Starting with solutions of the type (\ref{flatbrane}) one can easily
construct braneworld geometries by using the standard cut-and-paste
procedure.  For instance, the metric
\begin{eqnarray}
ds^2=dy^2 + e^{-2|y|/l} g_{\mu\nu}(x) dx^\mu dx^\nu
\label{rs}
\end{eqnarray}
represents a braneworld geometry with $\mathbb{Z}_2$ symmetry with
respect to the location of the brane ($y=0$). In $D=5$,
if we take $g_{\mu\nu}$ to be the 4-dimensional Schwarzschild metric
we reproduce the black string geometry~\cite{Chamblin:1999by}.
Instead, if we take $g_{\mu\nu}$ to be a pp-wave we have a
5-dimensional pp-wave travelling parallel to the brane~\cite{Chamblin:1999cj}.

This technique can be further generalized to the stacking of any
Einstein manifold.  Using the ansatz for the $D$-dimensional
metric\footnote{It is interesting to note that by using additional
  freedom in the way in which a $(D-1)$-dimensional geometry is
  embedded into a $D$-dimensional Einstein manifold one can locally and
  isometrically embed whatever metric \cite{Anderson:2001qt}.} 
\begin{eqnarray}
ds^2=dy^2 + e^{-2A(y)} g_{\mu\nu}(x) dx^\mu dx^\nu\,,
\label{brane}
\end{eqnarray}
the Einstein equations~(\ref{Einstein}) can be split into
\begin{eqnarray}
G_{\mu \nu }(x)=-C_{2}(y)g_{\mu \nu }(x)\,,
\label{Einsteinmunu}
\end{eqnarray}
which correspond to the $\mu\nu$ components, and
\begin{eqnarray}
C_{3}(y)+C_{4}(y) R(x)=0,
\label{Einsteinyy}
\end{eqnarray}
corresponding to the $yy$ component, where $R=g^{\mu\nu}R_{\mu\nu}$.
The coefficients are given by
\begin{eqnarray}
C_{2}(y) = &&\hspace{-2mm}
\left[\Lambda_{D}-(D-2)\left(A''-{(D-1) \over 2} A'^{2}\right)
\right]e^{-2A}, \\
C_{3}(y) = &&\hspace{-2mm}
\Lambda_{D}+{(D-1)(D-2) \over 2} A'^{2}, \\
C_{4}(y) = &&\hspace{-2mm}
-{1 \over 2}e^{2A}.
\end{eqnarray}
By differentiating equation (\ref{Einsteinyy}) with respect to any brane 
coordinate $x^\mu$ we obtain
\begin{eqnarray}
C_{4}(y) R(x)_{,\mu }=0,
\end{eqnarray}
so that the Ricci scalar has to be constant. For convenience we write 
this constant as
\begin{eqnarray}
R=\pm {(D-1)(D-2) \over  L^2}.
\end{eqnarray}
From Eq.~(\ref{Einsteinyy}) and for a negative cosmological constant
in the bulk, we have an equation for $A(y)$ of the form
\begin{eqnarray}
{(D-1)(D-2) \over 2}
\left(-{1 \over l^2} +A'^{2} \mp {1 \over L^2}e^{2A}
\right)=0\,.
\label{factorA}
\end{eqnarray}
Then Eq.~(\ref{Einsteinmunu}) becomes
\begin{eqnarray}
G_{\mu \nu }(x)=\mp {(D-2)(D-3) \over 2L^2}g_{\mu \nu }(x)\,,
\label{Einsteinmunu2}
\end{eqnarray}
which shows that $g_{\mu \nu }(x)$ must be a $(D-1)$-dimensional Einstein
metric. Equation~(\ref{factorA}) can be easily
solved.  When $L\rightarrow \infty$
we have $A(y)=y/l+b$, i.e., the solution in Eq.~(\ref{flatbrane})
(the constant $b$ is irrelevant for the geometry).
For $L \neq 0$ and defining $A=-\ln B$ we have
\begin{eqnarray}
B'^2={B^2 \over l^2} \pm {1 \over L^2}.
\label{eqforB}
\end{eqnarray}
The solution corresponding to the plus sign
(positively curved brane) is
\begin{eqnarray}
ds^2=dy^2 + \left({l \over L}\right)^2 \sinh^2\left({y-y_0 \over l}\right)
g_{\mu\nu}(x) dx^\mu dx^\nu,
\label{sol1}
\end{eqnarray}
and that corresponding to the minus sign (negatively curved brane) is
\begin{eqnarray}
ds^2=dy^2 + \left({l \over L}\right)^2 \cosh^2\left({y-y_0 \over l}\right)
g_{\mu\nu}(x) dx^\mu dx^\nu.
\label{sol2}
\end{eqnarray}
Again one can start from these bulk solutions to build braneworld models.
The braneworld models with maximally symmetric branes, 4-dimensional
de Sitter and anti-de Sitter branes~\cite{dewolfe}, are the simplest
illustration of this procedure. What we have shown here is that by using the
same warp factors one can have not only maximally symmetric branes
but {\em any} brane of the Einstein type.

For the case of a positive cosmological
constant in the bulk, $B(y)$ satisfies
\begin{eqnarray}
B'^2=-{B^2 \over l^2} \pm {1 \over L^2},
\end{eqnarray}
which has solutions only for the plus sign
(positively curved sections):
\begin{eqnarray}
B(y)={l \over L}\sin\left({y-y_0 \over l}\right),
\;\;\;\; {\rm or} \;\;\;\;
B(y)={l \over L}\cos\left({y-y_0 \over l}\right).
\end{eqnarray}
The case with zero bulk cosmological constant can be obtained
by taking the limit $l\rightarrow \infty$,
\begin{eqnarray}
B(y)={y-y_0 \over L}.
\end{eqnarray}

For completeness, it is interesting to observe
that this technique can also be used to generate bulk solutions
with positive cosmological constant and arbitrarily curved
sections. The result in this case
differs from the previous one in the following respect: given any Euclidean
metric $g_{\mu\nu}(x)$ such that
$R_{\mu\nu}=\pm [(D-2)/L^2]g_{\mu\nu}(x)$, then
\begin{eqnarray}
ds^2=-dt^2 + B(t)^2 g_{\mu\nu}(x) dx^\mu dx^\nu,
\end{eqnarray}
with
\begin{eqnarray}
&&B(t)=e^{2t/l} \;\;\;\;{\rm for}\;\;\;\; L\rightarrow \infty\,, \\
&&B(t)={l \over L}\cosh\left({y-y_0 \over l}\right)
\;\;\;\;{\rm for \;\; the \;\; positive \;\; sign}\,, \\
&&B(t)={l \over L}\sinh\left({y-y_0 \over l}\right)
\;\;\;\;{\rm for \;\; the \;\; negative \;\; sign}\,,
\end{eqnarray}
are Lorentzian $D$-dimensional Einstein manifolds of positive curvature.
This might be of interest in standard cosmology.
One can deduce this result easily from the above solutions for negative bulk cosmological constant via the substitutions:
$l\rightarrow -il$, $t\rightarrow it$ and $L\rightarrow -iL$.

%---------------------------------------------------------------------
\section{Einstein-Gauss-Bonnet gravity}
\label{S:gauss-bonnet}
%---------------------------------------------------------------------

Our central aim is to investigate the stacking
technique in the simplest modification of Einstein gravity for
higher-dimensional spacetimes, the Einstein-Gauss-Bonnet theory.
From now on we will set $D=5$. The theory is defined by the action
\begin{eqnarray}
S={1 \over 2 \kappa_5^2 }
\int dx^5 \sqrt{-^{(5)\!}g} \left[\,^{(5)\!}R-2\Lambda_5 +\alpha\,^{(5)\!} L_{GB} \right],
\end{eqnarray}
where
\begin{eqnarray}
^{(5)\!}L_{GB}=\,^{(5)\!}R^{ABCD}\,^{(5)\!}R_{ABCD}-4\,^{(5)\!}R^{AB}\,^{(5)\!}R_{AB}+\,^{(5)\!}R^{2},
\end{eqnarray}
and $\alpha$ is the coupling constant.
The field equations associated with the Einstein-Gauss-Bonnet action are
\begin{eqnarray}
^{(5)\!}G_{AB}=-\Lambda _{5}\,^{(5)\!}g_{AB}-\alpha\,^{(5)\!}H_{AB}\,,
\label{modfeq}
\end{eqnarray}
with
\begin{eqnarray}
^{(5)\!}H_{AB}=2\,^{(5)\!}R_{ACDE}\,^{(5)\!} R_{B}{}^{CDE}-4\,^{(5)\!}R_{ACBD}\,^{(5)\!}R^{CD}
-4\,^{(5)\!}R_{AC} \,^{(5)\!} R_{B}{}^{C}+2\,^{(5)\!}R\,^{(5)\!}R_{AB}-\frac{1}{2}\,^{(5)\!}g_{AB}\,^{(5)\!}L_{GB}.
\end{eqnarray}
This tensor is divergence free, $\nabla^A\,^{(5)\!}H_{AB}=0$, and so it
can be considered as some sort of source in the Einstein equations.
The existence of this term has dramatic consequences
for the stacking procedure.

%---------------------------------------------------------------------
\subsection{Stacking with a Gauss-Bonnet term}
%---------------------------------------------------------------------

We consider a metric of the form given in Eq.~(\ref{brane}), and
introduce it into the Einstein-Gauss-Bonnet equations. By doing this, we obtain
a set of equations playing the role of the effective $4$-dimensional Einstein equations
[the $\mu\nu$-components of Eq.~(\ref{modfeq})],
\begin{eqnarray}
C_{1}(y) G_{\mu \nu }(x)=-C_{2}(y)g_{\mu \nu }(x)
%-C_{3}(y) L_{GB}(x) g_{\mu \nu }(x),
\label{GBmunu}
\end{eqnarray}
and an additional condition [the $yy$-component of Eq.~(\ref{modfeq})],
\begin{eqnarray}
C_{3}(y)+C_{4}(y) R(x) + C_{5}(y) L_{GB}(x)=0,
\label{GByy}
\end{eqnarray}
where
\begin{eqnarray}
C_{1}(y) = &&\hspace{-2mm}
1+4\alpha \left(A''-A'^{2}\right), \\
C_{2}(y) = &&\hspace{-2mm}
\left[\Lambda_{5}-3\left(A''-2A'^{2}\right)+
12 \alpha A'^{2}\left(A''-A'^2\right)
\right]e^{-2A}, \\
C_{3}(y) = &&\hspace{-2mm}
\Lambda_{5}+6 A'^{2}(1-2 \alpha A'^{2}), \\
C_{4}(y) = &&\hspace{-2mm}
-\left({1 \over 2}-2 \alpha A'^{2} \right) e^{2A}, \\
C_{5}(y) = &&\hspace{-2mm}
-{\alpha \over 2} e^{4A}.
\end{eqnarray}
By dividing equation (\ref{GByy}) by $C_5(y)$ and differentiating 
with respect to $y$, we obtain 
\begin{eqnarray}
\left(C_{3}(y) \over C_{5}(y)\right)' - 
\left(C_{4}(y) \over C_{5}(y)\right)'
{}^{(4)\! }R(x)=0.
\label{condition1}
\end{eqnarray}
Now, differentiating it with respect to any brane coordinate $x$,
we arrive at
\begin{eqnarray}
\left[C_{4}(y) \over C_{5}(y)\right]'
\,R(x)_{,\mu }=0,
\label{condition}
\end{eqnarray}
so that either $R(x)_{,\mu }=0$ or
$(C_{4}/C_{5})'=0$.
We call the first case physical because it has a well defined
limit when $\alpha$ tends to zero, the Einsteinian limit.
Einstein-Gauss-Bonnet theory is ambiguous
from a dynamical point of view. The Lagrangian is quartic in
first derivatives of the metric and thus the same initial data can give
rise to different evolutions~\cite{zanelli}. To resolve this
ambiguity, one possibility is to choose as the physical branch
that approaching proper general relativity in the limit
$\alpha =0$~\cite{MenaMarugan:ji}. The second case, without an Einsteinian 
limit, we call ``purely" Gauss-Bonnet.

%---------------------------------------------------------------------
\subsection{Physical case: $R_{,\mu}=0$}
%---------------------------------------------------------------------

We write the constant Ricci scalar as
\begin{eqnarray}
R=\pm {12 \over L^2}.
\end{eqnarray}
By Eq.~(\ref{GByy}), $L_{GB}$ is also a constant, $L_{GB}=S_1$, and then
Eq.~(\ref{GByy}) gives
\begin{eqnarray}
6 A'^{2}(1-2 \alpha A'^{2})
-\left({1 \over 2}-2 \alpha A'^{2} \right) e^{2A}
\left(\pm {12 \over L^2}\right)
-{\alpha \over 2} e^{4A}S_1+\Lambda_{5}=0\,,
\label{eq3}
\end{eqnarray}
or re-arranging,
\begin{eqnarray}
2 \alpha A'^{4}-
\left(1 \pm  {4 \alpha \over L^2} \right) A'^{2}
+\left(-{\Lambda_5 \over 6} \pm {1 \over L^2}e^{2A}
+{\alpha \over 12}e^{4A}S_1 \right)=0\,.
\label{warpfactor}
\end{eqnarray}
Multiplying Eq.~(\ref{eq3}) by $e^{-4A}$ and differentiating we see
that equations~(\ref{GBmunu}) can be re-expressed as
proper Einstein equations for the 4-dimensional geometry,
\begin{eqnarray}
G_{\mu \nu }(x)=\mp {3 \over L^2} g_{\mu \nu }(x).
\end{eqnarray}
In summary, {\em for a metric of the form Eq.~(\ref{brane}) to be a
  solution of Einstein-Gauss-Bonnet theory in five dimensions, the
  4-dimensional metric $g_{\mu \nu }$ must be an Einstein metric.}
Moreover, and this is the important point, {\em it must have a
  constant Gauss-Bonnet term, i.e.} $L_{GB}=$constant.  These
conditions imply that the Krestschmann scalar, or equivalently in this
case the square of the 4-dimensional Weyl curvature, has to be
constant.  Their values in terms of the constants previously
introduced are
\begin{eqnarray}
R^{\mu\nu\gamma\sigma} R_{\mu\nu\gamma\sigma}
= S_1\,, ~~~~~
C^{\mu\nu\gamma\sigma} C_{\mu\nu\gamma\sigma}
= S_1 -\frac{24}{L^4}\,.
\end{eqnarray}
This condition precludes the existence of stacking solutions
of the black-string type~\cite{Chamblin:1999by} and most of the solutions
of astrophysical interest. The trivial conformal nature of maximally symmetric
and pp-wave spacetimes allows them to be stacked to form 5-dimensional bulk
solutions and subsequently braneworld models.

For arbitrary values of the constant $S_1$ the solution of
Eq.~(\ref{warpfactor}) cannot be expressed in closed form. However,
for the particular value $S_1={24 / L^4}$, which is the relevant value
for stacking maximally symmetric metrics on the brane, closed-form
solutions can be given. For $0>\Lambda_{5}=-6/l^2$,
\begin{eqnarray}
 A'^{2}={1 \over 4 \alpha}
\left(1 \pm  {4 \alpha \over L^2} e^{2A} +\sigma
\sqrt{1-{8 \alpha \over l^2} }
\right)
\label{roots}
\end{eqnarray}
where $\sigma=\pm 1$.
This expression has a well defined limit for $\alpha \rightarrow 0$
only when $\sigma=-1$. For this case and $0\leq\alpha \leq l^2/8$,
we can define $\tilde l$ and $\tilde L$ as
\begin{eqnarray}
&&{1 \over {\tilde l}^2 }={1 \over 4 \alpha}\left(1-
\sqrt{1-{8 \alpha \over l^2} }\right), \\
&&{1 \over {\tilde L}^2 }={4 \alpha \over L^2}.
\end{eqnarray}
Then, substituting $A=-\ln B$, Eq.~(\ref{roots}) is seen to be
equivalent to Eq.~(\ref{eqforB}), i.e.,
\begin{eqnarray}
B'^2={B^2 \over {\tilde l}^2} \pm {1 \over {\tilde L}^2}.
\label{eqforBtilde}
\end{eqnarray}
Therefore the warp factors for the Einstein-Gauss-Bonnet theory
are formally equal to those discussed in Sec.~\ref{S:stackingtechnique}
but with re-defined values of the curvature of the stacking hypersurfaces
and the bulk curvature.

Solutions with $\alpha<0$ are completely equivalent\footnote{In the
  Gauss-Bonnet modification to General Relativity arising from string
  theory, $\alpha$ must be positive. From this viewpoint, these
  solutions are not physical.}.  Their only peculiarity is that
positively-curved and negatively-curved stacking hypersurfaces
interchange their associated warp factors.

The solutions for the case $\sigma=1$ are of purely
Einstein-Gauss-Bonnet type in the sense that they do not have a well
defined Einsteinian limit. Particularly interesting is the fact that
in this case the Gauss-Bonnet term can produce a negatively curved
bulk even in the absence of a bulk cosmological
constant~\cite{Meissner:2000dy}. Here we take the view that these
solutions are unphysical. However, a definitive analysis of their
physical nature should consider not their proximity to 5-dimensional
Einstein solutions, but their proximity to 4-dimensional Einstein
solutions in the reduced 4-dimensional theory that emerges once the
compactification (exotic or not) has taken place.  This is beyond the
scope of the present paper.

%---------------------------------------------------------------------
\subsection{Purely Gauss-Bonnet case}
%---------------------------------------------------------------------

This case is characterized by $(C_{4}/C_{5})'=0$, so that condition 
(\ref{condition}) is satisfied. Therefore,
\begin{eqnarray}
{C_{4}(y) \over C_{5}(y)}=
\left({1 \over \alpha}-4 A'^{2} \right) e^{-2A} = S_2,
\label{cond1}
\end{eqnarray}
where $S_2$ is a constant. From Eq.~(\ref{GByy}),
\begin{eqnarray}
{C_{3}(y) \over C_{5}(y)}+S_2 R(x) + L_{GB}(x)=0\,,
\end{eqnarray}
and therefore
\begin{eqnarray}
&&{C_{3}(y) \over C_{5}(y)}=
-{2 \over \alpha} e^{-4A}
\left[\Lambda_{5}+6 A'^{2}\left(1-2 \alpha A'^{2}\right) \right]=S_3, \label{cond2} \\
&&S_2 R(x) + L_{GB}(x)=S_3,
\end{eqnarray}
with $S_3$ another constant.
By Eqs.~(\ref{cond1}) and (\ref{cond2}), we deduce that
there are two cases:

\noindent
{1)} $S_2^2=-{2 \over 3}S_3\neq 0$, $\Lambda_{5}=-{3 / 4 \alpha}$.

Again, for $\alpha>0$ we can define constants $\bar l$ and $\bar L$ via
\begin{eqnarray}
{1 \over {\bar l}^2}= {1 \over 4\alpha}, \\
{1 \over {\bar L}^2}={|S_2| \over 4}.
\end{eqnarray}
Setting $A=-\ln B$ we arrive at an equation equivalent to Eq.~(\ref{eqforB}),
\begin{eqnarray}
B'^2={B^2 \over {\bar l}^2} +{\rm sign}(S_1) {1 \over {\bar L}^2},
\label{eqforBtilde2}
\end{eqnarray}
and therefore, to the same formal solutions for the warp factors.
For these warp factors, we can check that $C_1(y)=C_2(y)=0$ and
so we reach the intriguing result that there is a single equation
for the 4-dimensional metric
\begin{eqnarray}
S_2 R(x) + L_{GB}(x)= -{3 \over 2} S_2^2.
\label{cond3}
\end{eqnarray}

\noindent
{2)} $A(y)=$constant.

Without lost of generality we can
set $A(y)=0$, and then $S_2=1/\alpha$ and $S_3=2\Lambda_5/\alpha$.
In this case the reduced set of equations to solve is, by
Eqs.~(\ref{GBmunu}) and (\ref{cond2}),
\begin{eqnarray}
G_{\mu\nu}=-\Lambda_5g_{\mu\nu}, \;\;\;\;
L_{GB}=-{2\Lambda_5 \over \alpha}.
\end{eqnarray}
If moreover $\Lambda_5=0$, these equations do not depend
on $\alpha$ and therefore, this case does in fact have a
well defined Einsteinian limit.

%---------------------------------------------------------------------
%\section{Junction conditions}
%\label{S:junction}
%---------------------------------------------------------------------

%---------------------------------------------------------------------
\section{Discussion and conclusions}
%---------------------------------------------------------------------

We have seen that in the presence of a Gauss-Bonnet term  the generation
of 5-dimensional bulk solutions via simple stacking of 4-dimensional metrics
does not hold as may have been naively expected. Although the Gauss-Bonnet
term is a topological invariant in four dimensions, in the process of
reducing the system from five dimensions to four dimensions the dynamical
degrees of freedom of the Gauss-Bonnet term in five dimensions
leave their traces in the reduced theory (cf.~\cite{Banados:1996gi}).

Let us consider a different point of view.
In the simplest case, with zero bulk cosmological constant
and no warp factor ($A=0$), we substitute the ansatz
$ds^2=dy^2+g_{\mu\nu}(x)dx^\mu dx^\nu$ into
the 5-dimensional action,
\begin{eqnarray}
S={1 \over 2 \kappa_5^2 }
\int dx^5 \sqrt{-^{(5)\!}g} \left[\,^{(5)\!}R +\alpha\,^{(5)\!} L_{GB} \right],
\label{5dim}
\end{eqnarray}
and integrate over $y$ (one can
consider a cylindrical condition on $y$ to obtain a finite result.)
In this way one finds a reduced action in four dimensions of the 
Einstein-Hilbert form,
\begin{eqnarray}
S={1 \over 2 \tilde \kappa_5^2 }
\int dx^4 \sqrt{-g} R.
\label{4dim}
\end{eqnarray}
(The trivial additive constant left by the Gauss-Bonnet 
   term has not been written. The Gauss-Bonnet term has become 
   topological in the reduction process.)
This is the standard Kaluza-Klein reduction in the absence of the
electromagnetic part of the metric (in Kaluza-Klein terminology);
therefore we recover vacuum Einstein gravity in four
dimensions~\cite{Muller-Hoissen:1990vf}. In standard Kaluza-Klein
reduction, the 5-dimensional Gauss-Bonnet term modifies the way in
which the electromagnetic field interacts with
gravity~\cite{Soleng:1994ps}, but in the absence of this field, there
is no other higher-dimensional manifestation of the Gauss-Bonnet term.
However, what we have seen is that if one uses first the action
Eq.~(\ref{5dim}) to obtain the 5-dimensional equations of motion, and
then one specializes to the above metric ansatz, one does recover the
4-dimensional Einstein equations $R_{\mu\nu}=0$, but in addition, one
finds the condition $L_{GB}=0$.  This condition puts a strong
restriction on the allowed geometries. Actually, this condition is
equivalent to $C_{\mu\nu\rho\sigma}C^{\mu\nu\rho\sigma}=0$, so it only
permits the existence of conformally trivial 4-geometries of this
type. Therefore, in the presence of a Gauss-Bonnet term, the process
of dimensional reduction of the action and subsequent variation is not
equivalent to the process of first varying the 5-dimensional action
and then reducing dimensionally the resulting equations.

What we have seen in this paper is that this phenomenon is general and
shows up in braneworld (exotic) compactifications as well.  Our simple
ansatz does not allow any curvature singularity on the brane, in
particular ruling out black string type braneworlds.  In this sense
our result is in tune with the well known idea that Gauss-Bonnet
corrections to Einstein relativity might smooth out the singularities.
Trying to find non-trivial brane geometries would involve the
consideration of an electromagnetic and/ or a dilaton part for the
metric with their corresponding effects.

%----------------------------------------------------------------
\section*{Acknowledgments}
%----------------------------------------------------------------

We thank Cristiano Germani and Bruce Bassett for useful discussions 
and comments. CB is supported by the EC under contract HPMF-CT-2001-01203.
RM is supported by PPARC. CFS and FV are supported by the EPSRC.

%========================================================================
% When possible, the references have proper Spires citations attached.
% This is supposed to help the Spires staff in updating their database.
% Don't touch the commented ``citation = '' lines.
%========================================================================

%----------------------------------------------------------------

%----------------------------------------------------------------
\end{document}